# A comprehensive study of Game Theory applications for smart grids, demand side management programs, and transportation networks

## Ali Mohammadi[1] and Sanaz Rabinia[2]

[1]Department of Electrical Engineering, Louisiana State University, Baton Rouge

[2]Department of Computer Science, Louisiana State University, Baton Rouge

Email: a.mohammadi004@gmail.com

Email: srabin1@lsu.edu

*Abstract*- Game theory is a powerful analytical tool for modeling decision makers strategies, behaviors and interactions. A Decision maker's ac and decisions can benefit or negatively impact other decision makers interests. Game theory has been broadly used in economics, politics and engineering field. For example, game theory can model decision making procedure of different companies competing with each other to maximize their profit.

In this chapter, we present a brief introduction of game theory formulation and its applications. The focus of the chapter is noncooperative Stackelberg game model and its applications in solving power system related problems. These applications include but not limited to; expanding transmission network, improving power system reliability, containing market power in the electricity market, solving power system dispatch, executing demand response and allocating resource in a wireless system. Finally, this chapter elaborates on solving a game theory problem through an example.

*Keywords*- Game theory, Stackelberg game, Nash equilibrium, Utility function, Smart Grid, Demand response, demand-side management

# 1. Introduction

Operation of power system relays on various entities with different objectives, sometimes with conflicting goals, and game theory lends itself really well to formulate this decision-making space. Game theory is a powerful tool for modeling and understanding strategies that can maximize a player's benefit in the context of other players' strategies. In the following chapter, we discuss the applications of game theory in solving power system related problems [1].

- **Power market**

Electricity market plays a crucial role in the operation of every power system market. The two pillars of any electricity market are generation companies' production bids and large consumers' consumption offers.

Generation companies bid production blocks and associated prices to the market's independent operator. Then, the system operator clears the market by determining the accepted bids [2]. This is a bi-level problem that can be modeled as a multi-leader-followers game problem. Game theory can be also leveraged to a model participation of renewable power producers in the electricity market [3, 4].

- **Power system dispatch**

Reliable and cost-effective operation of a power system is the primary responsibility of power system operators. Ever increasing penetration level of interment power production resources challenges generation and consumption balance and endangers reliability of the grid operation [5]. The zero-sum two-stage game is used in [6] to model a robust power dispatch in a power system. Along the same line,[7] models preventive and corrective actions in the unit commitment problem with a three-stage dynamic game.

- **Transportation network application**

Game theory has been widely used for solving electric vehicle coordination and optimization problems. In this regards, [8, 9] leverage game theory to derive optimal charging schedules for a group of electric vehicles in the context of power system contstraints.

- **Power system control**

Power system control strategies are designed around handling disturbances in the power grid. A disturbance can occur in both generation and consumption sides. One of the most effective robust control tools for dealing with disturbance is the differential game model. In this regard[10] leverages the differential game model to control disturbances of production and load sides of the power system. Also, the coordination between tie-line scheduling and frequency control is modeled based on the concept of the cooperative game [11].

- **Transmission expansion planning**

Transmission expansion planning (TEP) problem is aimed at determining optimal configuration for expansion of electricity network while ensuring the network's reliability. Given the enormous size of demand and supply scenario sets, installing a new line can be complex. In this regards, authors in [12] provide an overview of game theory based methods to solve TEP.

### 2.1. Game Theory definition

Game theory allows for mathematical modeling of interactions between decision-makers in the case that players' decisions influence each other directly. In a non-cooperative game theory, decision-makers are expected to act rational and selfish because they need to maximize their own benefits. On the other hand, in the cooperative game decision makers compete with each other while collaborating to fulfill an external enforcement (e.g., contract) [13-15].

#### 2.1.1 Elements of a game

Each game includes the following elements:

**Player set:** Number of players, denoted by $N$
**Strategy set:** A set of actions for each player $i$ which is denoted by $A_i$. The whole game space is obtained by Cartesian product of players actions, i.e., $A = A_1 \times A_2 \times ... \times A_n$. Defining $a_i \in A_i$, then $(a_i, a_{-i})$ can be assigned to the player $i$. Where $a_i$ denotes the player $i$ strategy while $a_{-i}$ denotes other players' strategies.

**Utility functions**: The utility function of player $i$ denoted by $u_i(a_i, a_{-i})$ which measure the utility outcome for player $i$ based on its strategy and other players' strategies. In a game theory players are considered rational meaning they are looking for the best strategy to maximize their benefit (utility) function.

#### 2.1.2 Decision rules:

Decision makings are executed based on specific set of rules. The process that players follow for selecting their strategies is called decision rules. In the following, we present decision rules that players are obliged to follow.

- **Best response dynamic Rule:**

Each player selects the best strategy based on other players' strategies to maximize self-benefit.

- **Better Response Dynamic Rule:**

Each time that a player executes a strategy, it is expected that this new policy increases that player's utility in comparison with its previous decision. A strategy can be selected randomly which may not be necessarily the best strategy. Note there is a trade-off between the convergence and complexity. Specifically, improving response of dynamic models can slow down the convergence

#### 2.1.3. Scheduling decision rules

Players should make their decision based on a specific time pattern. In general, we have four types of scheduling decision rules as follows:

**Synchronous decision rule**: In this scheduling decision rules, at each time step all players make their decisions simultaneously.

**Round robin decision rule**: Players make their decision sequentially until a decision making round ends.

**Random decision rule**: At each time step, a player plays randomly.

**Asynchronous decision rule**: At each time a random set of players make their decision simultaneously. Note this is the most commonly used rule.

### 2.1.4. Nash equilibrium, existence, and uniqueness

In the context of game theory, Nash equilibrium defines a steady state that all players are already selected their best strategy to maximize their benefits.

Mathematically, Nash equilibrium $\{N, A_i, U_i\}$ is defined as a profile $s_i \in S$ of actions of player $i \in N$ that lead to

$$u_i(s_i^*, s_{-i}^*) \geq u_i(s_i, s_{-i}^*)$$

In other words, player $i$ cannot change the strategy from $s_i^*$ to $s_i$ and achieves a better utility. Therefore, all players prefer to stay at this equilibrium point. In general, a game can have zero, one or more Nash equilibrium points. The theorem below discusses the conditions for achieving the Nash equilibrium.

**Kakutani Fixed Point Theorem:** A strategic game $\{N, (A_i), (u_i)\}$ has a Nash equilibrium if, for all $i \in N$, the action set $A_i$ of player $i$ is a non-empty compact convex subset of a Euclidian space and the utility function $u_i$ is continuous and quasi-concave on $A_i$ [13].

This theorem guarantees that a problem has at least one Nash equilibrium point, but it doesn't prove the uniqueness of that point. Therefore, the game problem can have several Nash points. There are other theorems concerning the uniqueness of Nash equilibrium point. In this regard, a well-recognized theorem is a supermodular game that details the required conditions for achieving a unique Nash point.

**Supermodular game:** A game is supermodular in the case that a game space makes a lattice and all utility functions are supermodular as well.

Note, X is a lattice if:

$\forall a, b \in X: a \wedge b \in X, a \vee b \in X$ while $a \vee b = SuP(a, b), a \wedge b = inf(a, b)$ and $f: X \to \mathbb{R}$

In this lattice we can call a game supermodular if:

$\forall a, b \in X: f(a) + f(b) \leq f(a \wedge b) + f(a \vee b)$

Reference [16], simplifies this mathematically complex definition and presents sufficient conditions for a supermodularity of a game [16].

**Theorem:** A game can be identified as a supermodular game, if the action space of each player is a closed subset on a real space and response function satisfies the following constraint,

$$\forall j \neq i \in N: \frac{\partial^2 u_i(a)}{\partial a_i \partial a_j} \geq 0$$

These games enjoy important properties including:

1) There is at least one Nash equilibrium point for these games.
2) Under the decision rule, the best response and synchronous and asynchronous timing rules converge to the unique Nash equilibrium point.
3) If the best response of each player $BR_i(a)$ satisfies the following conditions, the supermodular game will have a unique Nash equilibrium:

- Uniqueness: $BR_i(\underline{a}) = \{b_i \in A_i: u_i(b_i, a_{-i}) \geq u_i(a_i, a_{-i})\}$ is a singleton for all a.
- Positivity: $\forall i \in N, \underline{a} \in \underline{A}: BR_i(a) > 0$
- Scalability: $\forall i \in N, \alpha > 1, a \in A: \alpha BR_i(\underline{a}) > BR_i(\alpha \underline{a})$

So far, we have described the basics of game theory. The remainder of this chapter is dedicated to presenting Stackelberg Game which is commonly used to analyze and model power system related problems.

### 2.2. Stackelberg Game

The Stackelberg leadership game is a strategic game in which players are divided into to two groups; leader and follower. This game is based on two stages. At the first stage, the players of the leader group select their strategies and act on them. At the second stage, the follower group chooses and execute their strategies according to the leader group actions. This game is built on the assumption that leader players know the impact of their actions on other players. Moreover, there is no way for followers to disregard the leaders' actions.

Usually, the Stackelberg game defines the situation that a set of players have specific privileges that allow them to move first. Leaders must also be committed to followers. Once a leader makes a decision, it can no longer change the selected policy, because it is committed to that action.

#### 2.2.1: Subgame Perfect Nash Equilibrium (SPNE)

In order to solve the Stackelberg game, we have to find a Subgame Perfect Nash Equilibrium. Subgame Perfect Nash Equilibrium is the strategic vector that is the best strategy for each player with the assumption that the strategy is stable for other players. This further implies that each player in each sub-player plays according to a Nash equilibrium.

The game takes place in two steps. At the first step, the lead player calculates the best followers' response. In other words, leader calculates followers' responses to its actions. Then the leader predicts follower's responses and chooses the strategy and action to maximize its profit. At the second step, a follower observes strategies chosen by the leader. At the equilibrium state, follower selects the leader's expected value as a response to the leader action.

Therefore, to calculate the Subgame Perfect Nash Equilibrium, at first we need to calculate the best-response functions of the followers in backward induction. In order to better comprehend the procedure of finding the equilibrium point in a Stackelberg game, we have provided the following example.

**Example 2-1: Calculating Nash Equilibrium in a Stackelberg game**

Assume a Stackelberg game that includes two players (a leader and a follower) and utility function for each of leader or follower are given as

$$\Pi_1 = p(q_1, q_2) \cdot q_1 - C_1(q_1)$$
$$\Pi_2 = p(q_1, q_2) \cdot q_2 - C_2(q_2)$$

In which $q_1$, is the leader's strategy and $q_2$ is the follower's strategy, which determined by the leader's strategy, therefore $q_2$ is a function of $q_1$.

Functions $C_1(q_1)$ and $C_2(q_2)$ are also the cost functions of these two players, respectively. To ease the calculation burden, $C_1(q_1)$ and $C_2(q_2)$ are assumed depend on one player's strategy. Also, the function $p(q_1, q_2)$ is assumed the follow the below form:

$$p(q_1, q_2) = (a - b(q_1 + q_2))$$

The Stackelberg game structure requires the leader to determine its strategy first. As discussed before, the leader considers the impacts of its decisions on followers' decisions, since the followers' strategies affect the leader's utilization as well.

Therefore, first, the follower's best response to the leader movement must be calculated as a function of the leader's selected strategy. The best response to the follower's utility function is the value of $q_2$, with the assumption of knowing $q_1$, leads to a maximization of $\Pi_2$ (the follower's utility function). So with the assumption of knowing the leader's output, the output that maximizes the utility of the follower is calculated. So, to find the maximum value of $\Pi_2$ with respect to $q_2$, we calculate derivative of $\Pi_2$ with respect to $q_2$ and then equalize the derived equation to zero.

$$\frac{\partial \Pi_2}{\partial q_2} = \frac{\partial p(q_1,q_2)}{\partial q_2} \cdot q_2 + p(q_1,q_2) - \frac{\partial C_2(q_2)}{\partial q_2} = 0 \ (6\text{-}2)$$

The values of $q_2$ which satisfies the above equation are the best responses.

By substituting (6-2) in the above equation and solving it for $q_2$, the following equation achieves the best response of the follower to $q_1$:

$$q_2 = q_2(q_1) = \frac{a - bq_1 - \frac{\partial C_2(q_2)}{\partial q_2}}{2b}$$

Now, we evaluate the best response function. This function is calculated by considering the output of the follower as a function of the output of the leader.

The leader understands that selecting $q_1$, results in selecting $q_2$ by followers (based on equation (8-2)) as the best response to $q_1$. Therefore, the leader can calculate its maximum utility by substituting the above equation in its utility function. In this regard, we take the derivative of $\Pi_1$ with respect to $q_1$ and equalize that equation to zero.

$$\frac{\partial \Pi_1}{\partial q_1} = \frac{\partial p(q_1,q_2)}{\partial q_2} \cdot \frac{\partial q_2}{\partial q_1} \cdot q_1 + p(q_1, q_2(q_1)) - \frac{\partial C_1(q_1)}{\partial q_1} = 0 \quad (9-2)$$

By solving the above equation for $q_1$, we obtain the optimal move for the leader as,

$$q_1^* = \frac{a + \frac{\partial C_2(q_2)}{\partial q_2} - 2 \cdot \frac{\partial C_1(q_1)}{\partial q_1}}{2b} \quad (10-2)$$

This value is the leader's best response to the reaction of a follower at the equilibrium point. Now, we can find the value of the follower utility by substituting the above equation in the follower's reaction function that was previously derived (equation (8-2)):

$$q_2^* = \frac{a - b \cdot \frac{a + \frac{\partial C_2(q_2)}{\partial q_2} - 2 \cdot \frac{\partial C_1(q_1)}{\partial q_1}}{2b} - \frac{\partial C_2(q_2)}{\partial q_2}}{2b} \quad (11-2)$$

The resulting equation is:

$$q_2^* = \frac{a - 3 \cdot \frac{\partial C_2(q_2)}{\partial q_2} + 2 \cdot \frac{\partial C_1(q_1)}{\partial q_1}}{4b} \quad (12-2)$$

Therefore, the Nash equilibrium points of the game, are all possible $(q_1^*, q_2^*)$ responses (derived using equations (10-2) and (12-2)).

### 3. Conclusion:

Operation of power system is influenced by multiple entities with different and often conflicting interests. These entities compete in the pursuit of their self-interests. Game theory is a strong tool for modeling this competition between decision-making entities. In this book chapter, we presented properties of game theory in an effort to motivate power system researchers to use this powerful tool. We also briefly discussed multiple problems that can be modeled as a game. Finally, we presented a detailed discussion for one of the popular game theory problem formulations and provided an example to clarify its implementation.